\documentclass[prd,twocolumn,superscriptaddress,altaffilletter,showpacs,nofootinbib]{revtex4-1}

\usepackage{amssymb}
\usepackage{amsfonts}
\usepackage{amsthm}
\usepackage{graphicx}
\usepackage{color}
\usepackage{amsmath}
\usepackage{latexsym}
\usepackage{makeidx}\makeindex
\usepackage{titlesec}
\usepackage{fancyhdr}
\usepackage[utf8]{inputenc} 

\newcommand{\be}{\begin{equation}}
\newcommand{\ee}{\end{equation}}
\newcommand{\bea}{\begin{eqnarray}}
\newcommand{\eea}{\end{eqnarray}}

\begin{document}

\title{Thermoluminescent characteristics of LiF:Mg,Cu,P and CaSO$_4$:Dy for low dose measurement}

\author{S. Del Sol Fern\'andez} \email{susi2489@hotmail.com}\affiliation{CICATA-Legaria, Instituto Polit\'ecnico Nacional, M\'exico D.F., CP 11500, M\'exico.}

\author{Ricardo Garc\'ia-Salcedo} \email{rigarcias@ipn.mx}\affiliation{CICATA-Legaria, Instituto Polit\'ecnico Nacional, M\'exico D.F., CP 11500, M\'exico.}

\author{J. Guzm\'an Mendoza} \email{joguzman@ipn.mx}\affiliation{CICATA-Legaria, Instituto Polit\'ecnico Nacional, M\'exico D.F., CP 11500, M\'exico.}

\author{D. S\'anchez-Guzm\'an}\email{dsanchez@ipn.mx}\affiliation{CICATA-Legaria, Instituto Polit\'ecnico Nacional, M\'exico D.F., CP 11500, M\'exico.}

\author{G. Ram\'irez Rodr\'iguez} \affiliation{Hospital Ju\'arez de M\'exico, 07760, M\'exico D. F.}

\author{E. Gaona} \affiliation{Universidad Aut\'onoma Metropolitana-Xochimilco, 04960, M\'exico, D.F.}

\author{T. Rivera Montalvo} \email{trivera@hotmail.com}\affiliation{CICATA-Legaria, Instituto Polit\'ecnico Nacional, M\'exico D.F., CP 11500, M\'exico.}

\date{\today}

\begin{abstract}
Thermoluminescence (TL) characteristics for LiF:Mg,Cu,P, and CaSO$_4$:Dy under the homogeneous field of X-ray beams of diagnostic irradiation and its verification using thermoluminescence dosimetry is presented. The irradiation were performed utilizing a conventional X-ray equipment installed at the Hospital Ju\'arez Norte of M\'exico. Different thermoluminescence characteristics of two material were studied, such as batch homogeneity, glow curve, linearity, detection threshold, reproducibility, relative sensitivity and fading. Materials were calibrated in terms of absorbed dose to the standard calibration distance and they were positioned in a generic phantom. The dose analysis, verification and comparison with the measurements obtained by the TLD-100 were performed. Results indicate that the dosimetric peak appears at 202$^o$C and 277.5$^o$C for LiF:Mg,Cu,P and CaSO$_4$:Dy, respectively. TL response as a function of X-ray dose showed a linearity behavior in the very low dose range for all materials. However, the TLD-100 is not accurate for measurements below $4mGy$. CaSO$_4$:Dy is 80\% more sensitive than TLD-100 and it show the lowest detection threshold, whereas LiF:Mg,Cu,P is 60\% more sensitive than TLD-100. All material showed very good repeatability. Fading for a period of one month at room temperature showed low fading LiF:Mg,Cu,P, medium and high for TLD-100 and CaSO$_4$:Dy. The results suggest that CaSO$_4$:Dy and LiF:Mg,Cu,P are suitable for measurements at low doses used in radiodiagnostic.
\end{abstract}


\maketitle

\section{Introduction}

Medical applications of ionizing radiation are the most important sources irradiation of the population. Ionizing radiation is used in medicine in three areas: nuclear medicine, radiotherapy and diagnostic radiology, the latter use X-ray equipment to obtain images of the inside of the patient's body.

Dosimetric investigations in diagnostic radiology have been increasing in importance in the last two decades. The most widely used technique in radiation dosimetry is thermoluminescence (TL).

Several types of thermoluminescent dosimeters (TLD) are commercially available for a wide range of applications: personnel, environment and medical dosimetry, nuclear accidents, etc.

Lithium fluoride doped with magnesium and titanium, known commercially as TLD-100, is still the most commonly used radiation dosimeter. It has become popular because of several properties, such as tissue equivalence, relative low fading and the possibility to manufacture the material with acceptable reproducibility \cite{Dewerd1999}, \cite{Guimaraes2003}, \cite{IAEA1995}, \cite{Horowitz1984}. 

TLD-100 has some features that do not entirely suitable for use in low dose X-ray such as low sensitivity (which is why it is necessary to calibrate every use), poor detection threshold and disagreement in several reports about the fading \cite{DeWerd1979}, \cite{Oberhofer1981}, \cite{McKeever1985}. In this work this material was used only as a reference dosimeter. 

Other material with nearly tissue equivalence is lithium fluoride doped with magnesium, copper, and phosphorus (LiF:Mg,Cu,P) \cite{Bilski1995} and \cite{Budzanowski2007}. It has several important advantages compared to TLD-100 such as higher sensitivity, low fading, good detection threshold; however, this has not yet been proposed for routinely use for dosimetric applications as the TLD-100.

On the other hand, there are materials over-respond due to their higher effective atomic number $Z$. Thus, they have higher sensitivity and are characterized as non-tissue equivalent materials. This materials are calcium sulfate (CaSO$_4$) and calcium fluoride (CaF$_2$) among others and are used for environmental monitoring. 

For environmental monitoring of conventional medical installations, dosimeters should be placed at least a month, so its fading in this period of time should be the minimum. Calcium fluoride presents a rapid fading in a short period of time \cite{Sukis1971}, \cite{Furetta1983}, therefore this materials are not suitable for routine monitoring of low dose X-ray.

Calcium sulfate has advantages for environmental TLD. It can easily be prepared, is $\approx$ 30 times as sensitive as TLD-100, and exhibits considerably less fading than calcium fluoride \cite{Becker1972}. However this characteristics have been realized for indoor dosimetry of gamma radiation areas in around of the nuclear installations in the worldwide \cite{Vohra1980}, \cite{Benkrid1992}, \cite{Takale2014}. But not there are measurements of environmental X radiation delivered in medical installations.

The aim of this study is to determine the dosimetric characteristic of two different thermoluminescent materials LiF: Mg, Cu, P and pellets synthesized in Mexico based on CaSO$_4$:Dy, and compared them with TLD-100 for low dose X-rays used in diagnostic radiology in similar conditions or close to real working conditions which may reduce the uncertainties associated with commercial TLDs.

\section{Materials and Methods}

\subsection{Materials} \label{2.1}

The materials and equipment used are listed below: 

\begin{itemize}
    \item TL dosimeters
        \begin{itemize}
        \item 18 chips of TLD-100 of $3 \times 3 mm$ and $1 mm$ thick.  (Harshaw Chemical Company, Solon, OH, USA).
        \item 20 discs of $3mm$ diameter and $1mm$ thick of CaSO$_4$:Dy (CICATA-Legaria, IPN).
        \item 16 discs of $4.5mm$ diameter and $1mm$ thick of LiF:Mg,Cu,P (MIKROLAB s.c., Poland).
        \end{itemize}
    \item Harshaw Thermo Scientific TLD Reader (Model 3500).
    \item Harshaw Thermolyne heating muffle (Model 1400).
    \item Radcal Accu-Gold+ multisensor (Model AGMS-D).
    \item Conventional X-ray equipment (CMR model MRH-II E GMX 325AF SBV-1 with Rotating Anode X-ray tube).
    
\end{itemize}

\subsection{Methodology} \label{2.2}
Dosimeters received a standard annealing treatment before exposure to radiation. Depending on the type of material, thermal annealing schemes were: $240^o$C for 10 minutes to LiF:Mg,Cu,P and $300^o$C for 30 minutes to CaSO$_4$:Dy and $400^o$C for 1 hour followed by $100^o$C for 2 hours to TLD-100. The method of slow cooling inside the muffle was used to reach room temperature for all cases.

The radiation was performed with a conventional X-ray equipment of radiology area at Hospital M\'exico specifically designed for general radiographic procedures.

The readings of the TL materials are performed in a reader. The reading cycles were varied depending on the material as shown in Table \ref{T1}. In order to eliminate the contribution by triboluminescence all readings were performed in an atmosphere of high purity N$_2$.

\begin{table}[h] 
\centering
\begin{tabular}{| c | c | c |}
\hline \hline
Parameters & LiF:Mg, Cu, P & LiF:Mg, Ti \\
TTP        &  $ $          & CaSO$_4$:Dy \\
\hline\hline 
Preheating temperature & $100^oC$ &  $50^oC$\\
\hline
Preheating time & $12s$ &  $5s$\\

\hline
Preheating speed & $8^oC/s$ & $10^oC/s$\\
\hline
Max. Heating temperature             & $240^oC$                & $350^oC$\\
\hline
Adquisition time & $20s$ &  $30s$\\
\hline
Annealing temperature & $240^oC$ &  $350^oC$\\
\hline\hline
\end{tabular}\caption{Reading parameters for TLD materials}\label{T1}
\end{table}

\subsubsection{Batch homogeneity}
Batches of the three materials already described were annealed with the parameters listed above according to each type of material. They were then irradiated with a conventional X-ray equipment at a dose of $5 mGy$ (considering their possible use for clinical dosimetry) for the TLD-100 and LiF:Mg,Cu,P whereas the CaSO$_4$:Dy was irradiated at a dose $1 mGy$ (for applications in environmental dosimetry and personal).

Immediately after irradiation, the materials were read under the parameters mentioned above, the TL emission values were recorded as $M_i$ with $i=1,2,3,...,N$. Later they were erased and re-read to determine the background reading or reading to a zero dose and these values were recorded as $M_{0i}$. The average values of these readings was calculated and expressed in terms of:
\begin{equation}
\Delta_{max}=\frac{(M-M_0)_{max}-(M-M_0)_{min}}{(M-M_0)_{min}} \times 100\%
\end{equation}
where $\Delta_{max}$ is the index of uniformity for given batch and must be less than or equal to 30\% in order to be considered acceptable values \cite{Furetta1998}. 

\subsubsection{Glow curve}
To study the TL curves, 6 dosimeters were used, which were previously annealed, they were protected from light and irradiated to $80 kVp$, $300 mA$, with an exposure time of $0.5 s$, which means a dose of $14.69 mGy$. We used a focus-surface distance (FSD) of $80 cm$ and a field of $10 \times 10 cm^2$. Readings were taken at 24 hours post irradiation. 

\begin{figure}[h]
\includegraphics[width=8cm]{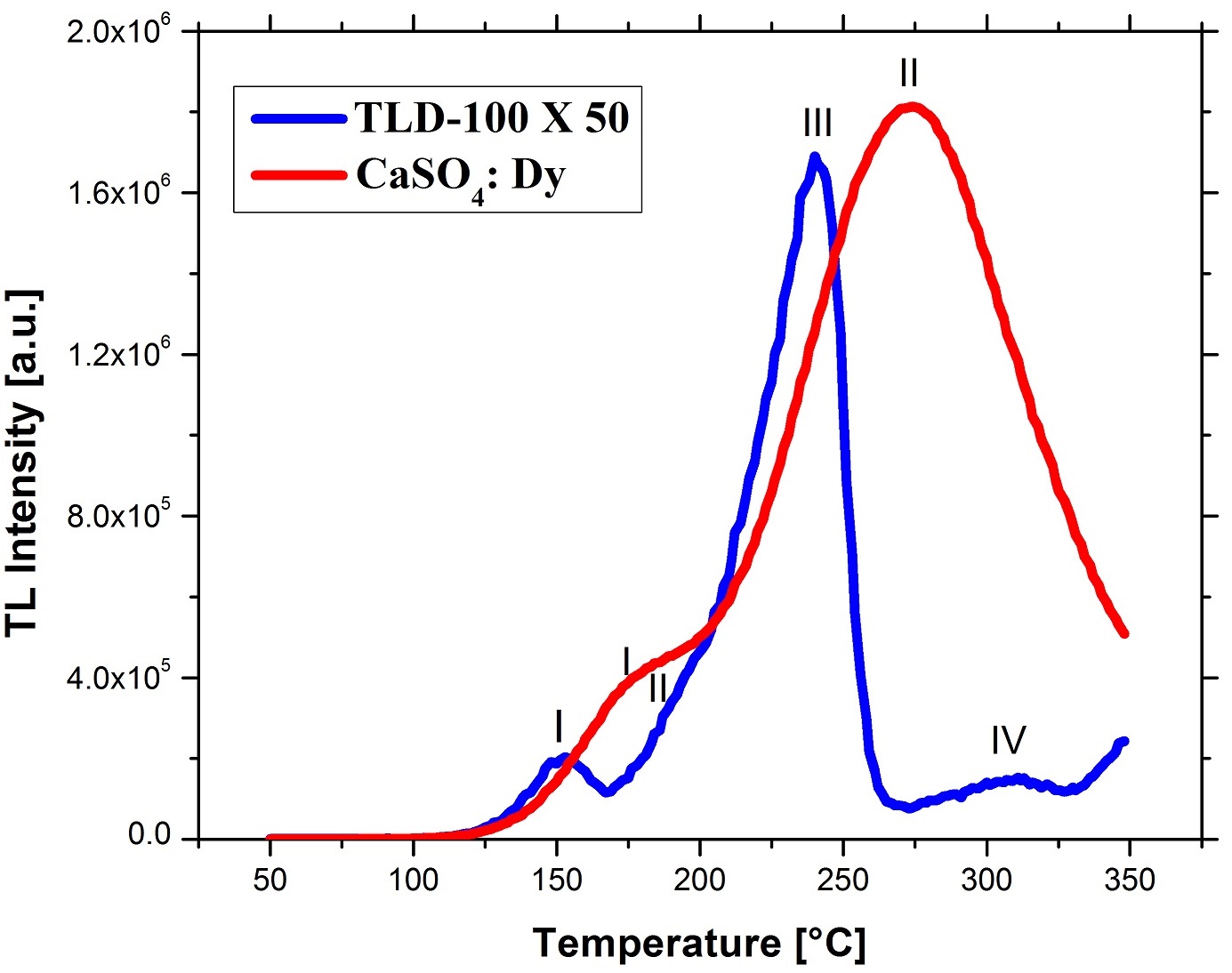}
\caption{Glow curve at a dose of $14.6 mGy$ for TLD-100 and CaSO$_4$:Dy. Note that the thermoluminescent intensity of TLD-100 is multiplied by 50.}
\label{F1a}
\end{figure}

\subsubsection{TL response as function of dose (Linearity)}
A solid state multisensor was used for calibration of TL materials in terms of absorbed dose. 16 dosimeters of each material were divided into groups of 4 and dosimeters were placed into capsules of latex, which are arranged adjacent to the AGMS. These were irradiated with the equipment of conventional radiology under the following parameters: $80 kVp$, $300 mA$ to a DFS $100cm$, with a size of $10 \times 10cm^2$ to different values of the product of current-time $18mAs$, $36mAs$, $75mAs$, $150mAs$, which gave the $1.75mGy$, $3.52mGy$, $7.29mGy$ and $14.69 mGy$ doses, respectively. Dosimeters were readed at 24 hours post-irradiation. In order to obtain the TL response as a function of the radiation dose for each of the materials, the TL intensities were plotted versus the obtained from AGMS in the range of doses studied.

\subsubsection{Detection threshold, $D_{LDL}$} 
The lower detection limit is defined as the lowest dose that can be detected with an acceptable confidence level \cite{Hirning1992}, which is defined as 3 times the standard deviation of the reading at zero dose, and is expressed in units of absorbed dose.

Four dosimeters of each material previously annealed as indicated in \ref{2.2}, were irradiated at a dose of $11.49 mGy$ for TLD-100 and $7.29mGy$ for LiF:Mg,Cu,P, and CaSO$_4$:Dy with the conventional X-ray equipment described above, they were readed at 24 hours after irradiation and an equal cycle is performed again. Detection thresholds for the three materials were calculated from the following expression:
\begin{equation}
D_{LDL}=3 \sigma_{BKG} \times \Phi_C, \nonumber
\end{equation}
where $\sigma_{BKG}$ is the standard deviation at zero dose and $\Phi_C=\frac{D}{M}$ is the calibration factor for determinated dose $D$.

\begin{figure}[h]
\includegraphics[width=8cm]{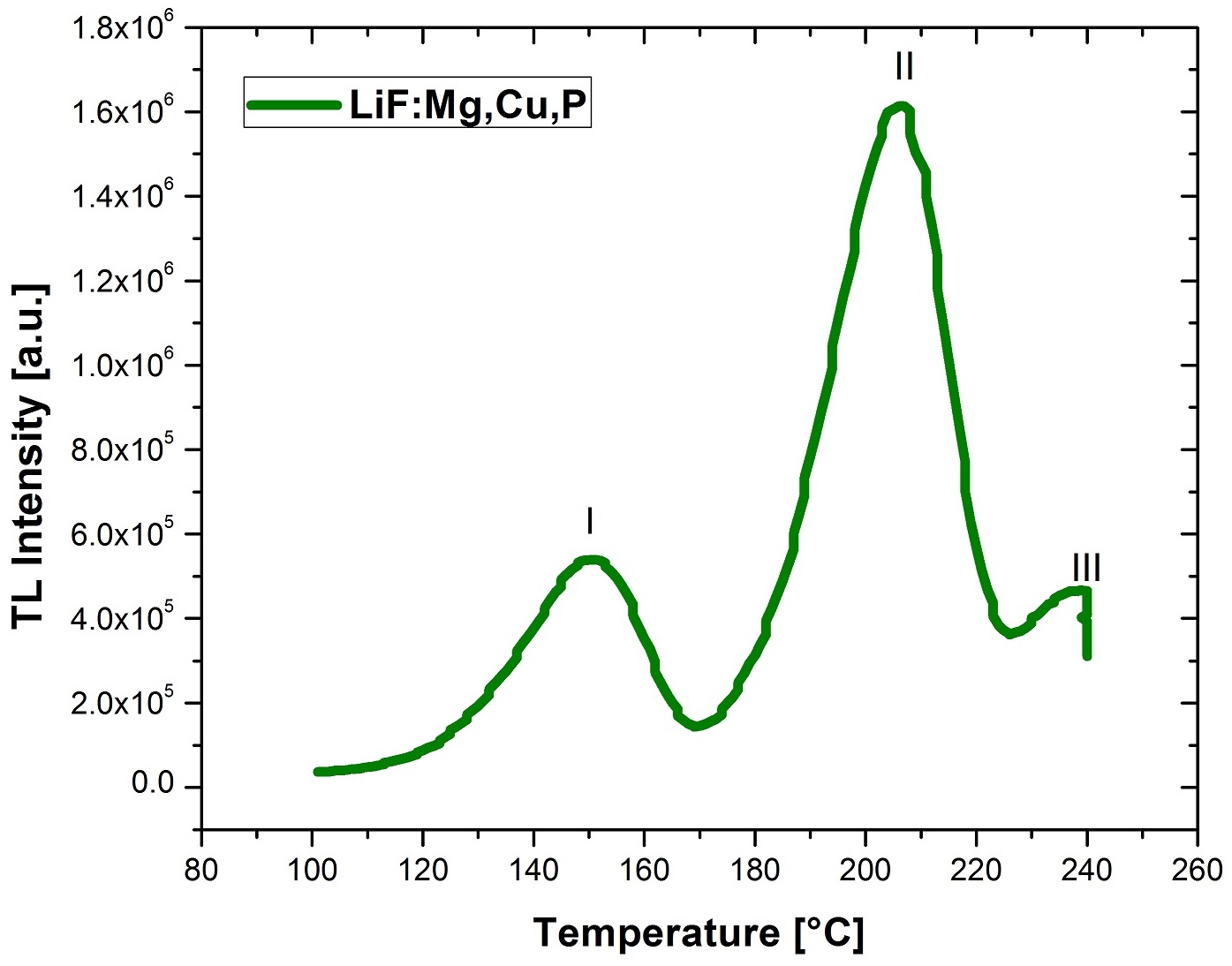}
\caption{Glow curve at a dose of $14.6 mGy$ for LiF:Mg,Cu,P.}
\label{F1b}
\end{figure}

\subsubsection{Repeatability} 
In order to study the repeatability of the material at low doses, a total of six dosimeters, two of each type were used. The test was performed for ten consecutive cycles, i.e., thermal annealing treatment, irradiation and reading with the same conditions for each annealing cycle. Annealing was conducted according to the conditions mentioned in Section \ref{2.2}, the irradiation was performed at a dose of 5mGy and readings were made 24 hours post-irradiation using the same parameters mentioned in section \ref{2.2}. The repeatability of the TL response as a function of absorbed dose was calculated with the following expression:
\begin{equation}
R=\frac{100 \sigma}{\bar{x}} \leq 7.5\%, \label{RR}
\end{equation}
where $\sigma$ is the standard deviation and $\bar{x}$ is the average of all readings during the 10 cycles.

\subsubsection{Relative sensitivity}
The relative sensitivity of each material was determinated by comparing the mean values obtained in the reproducibility test, considering TLD-100 as a reference.

\subsubsection{Fading at room temperature}
Fading of dosimeters as a function of time was studied. To do this, 12 dosimeters each type of material were used, previously annealed, then they were irradiated at a dose of $14 mGy$ and they kept it stored all the time at a temperature of $20^o$C. Readings were taken at the following post-irradiation time: $3h$, $24h$, $48h$, $120h$, $168h$, $288h$, $720h$ (1 month).

\section{Results and discussion}

In Figures \ref{F1a} and \ref{F1b} are shown TL glow curves obtained for the three different materials at low doses of X-rays used in the field of radiology.

The dosimeter LiF:Mg,Ti has a spectrum with four peaks centered at temperatures of $155^o$C, $192^o$C, $243^o$C and $305^o$C; CaSO$_4$:Dy has two peaks centered at $179^o$C and $277^o$C. Finally, LiF:Mg,Cu,P presented three peaks that appeared at $148^o$C, $202^o$C and $237^o$C. Dosimetric peaks for each material are centered to the following temperatures $T=243^o$C, $T=277^o$C and $T=236^o$C, respectively.

In Table \ref{T2} shows the values of TL response of different materials for the range of $1.76mGy$ to $14.69mGy$ during calibration or response in a dose dependent. It was obtained a variation in the relative standard deviation of the TL readings from $0.09$ to $0.32$ for the TLD-100, $0.02$ to $0.22$ for CaSO$_4$:Dy and $0.03$ to $0.2$ for LiF:Mg,Cu,P. It was observed that the standard deviation was very high at below $4mGy$ dose mainly for the TLD-100 ($32\%$). For this reason, many reports also analyze just over $5 mGy$ doses \cite{Livingstone2009}, \cite{Maia2010}. 

\begin{widetext}

\begin{table} [h]
\centering
\begin{tabular}{| c | c | c | c | c |}
\hline \hline
Exposition time & AGMS  & TLD-100 & CaSO$_4$:Dy & LiF:Mg,Cu,P \\

[s] & [mGy] & [nC] & [nC] & [nC] \\
\hline\hline 
$0.06$ & $1.758$ & $\bar{x}=42.9, \sigma=32\%$ & $\bar{x}=1201.2, \sigma=22\%$ & $\bar{x}=1127.0, \sigma=8\%$ \\
\hline
$0.12$ & $3.521$ & $\bar{x}=68.6, \sigma=20\%$ & $\bar{x}=2461.5, \sigma=16\%$ & $\bar{x}=2314.5, \sigma=5\%$ \\
\hline
$0.25$ & $7.291$ & $\bar{x}=115.4, \sigma=10\%$ & $\bar{x}=5064.0, \sigma=17\%$ & $\bar{x}=4727.5, \sigma=3\%$ \\
\hline
$0.5$ & $14.694$ & $\bar{x}=230.1, \sigma=9\%$ & $\bar{x}=10395.0, \sigma=2\%$ & $\bar{x}=9113.0, \sigma=20\%$ \\
\hline\hline
\end{tabular}\caption{Values obtained during calibration of the materials to low dose X-rays.}\label{T2}
\end{table}

\begin{table}[h] \centering
\begin{tabular}{| c | c | c |c |}
\hline \hline
TL material & Repeatability  & Detection threshold [$\mu Gy$] & Relative Sensitivity \\
\hline\hline 
TLD-100 & $3.18$ &  $160$ & $1$  \\
\hline 
CaSO$_4$:Dy & $3.20$ &  $6$ & $82.3$  \\
\hline 
LiF:Mg,Cu,P & $4.00$ &  $12$ & $66.3$  \\
\hline\hline
\end{tabular}\caption{Repeatability, detection threshold and relative sensitivity of different TLDs in a diagnostic X-ray beam.} \label{T3}
\end{table}

\end{widetext}

\begin{figure}[h]
\includegraphics[width=7cm]{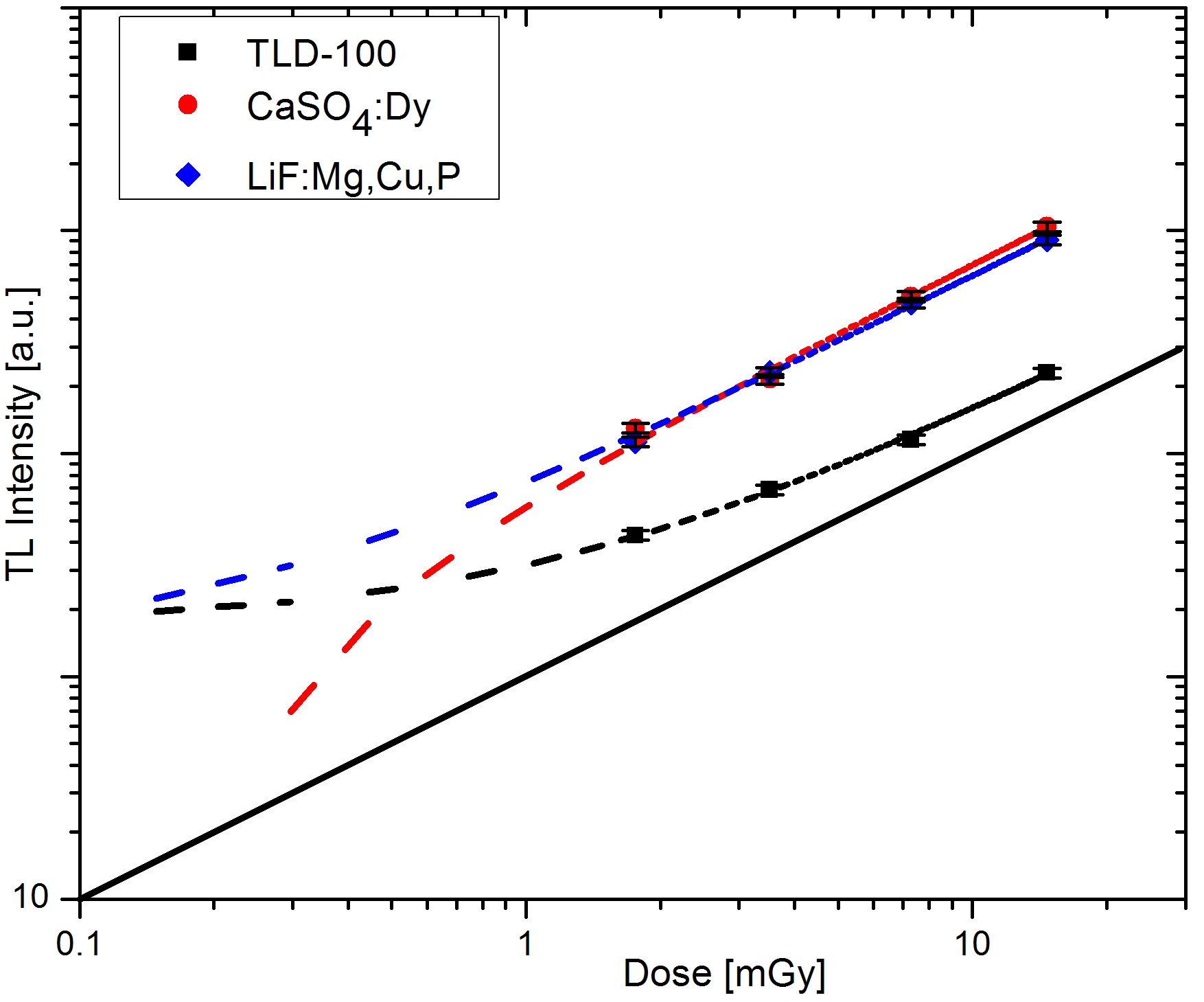}
\caption{TL response of materials as a function of dose.}
\label{FA}
\end{figure}

Figure \ref{FA} shows the the dose-response curve for an $80 kVp$ X-ray in the low dose in log-log scale. A linear plot in the log- log scale with the slope equal to 1 indicates a linear dose response. The error bars in the graph corresponds to $5$\%.  Non-linearity, as reported by some authors \cite{Cameron1968}, \cite{Watson1970} was observed for the TLD-100 below $4mGy$. These findings are important and should be made available to physiologist and occupationally exposed using TLD-100  for the monitoring of low and very low doses.

The values for repeatability test for each type of material followed by 10 cycles of irradiation, Eq. (\ref{RR}), are presented in Table \ref{T3}. Also in Table \ref{T3} it is shown the detection threshold and relative sensitivity for the TLD-100, LiF:Mg,Cu,P and CaSO$_4$:Dy. 

\begin{figure}[h]
\includegraphics[width=7cm]{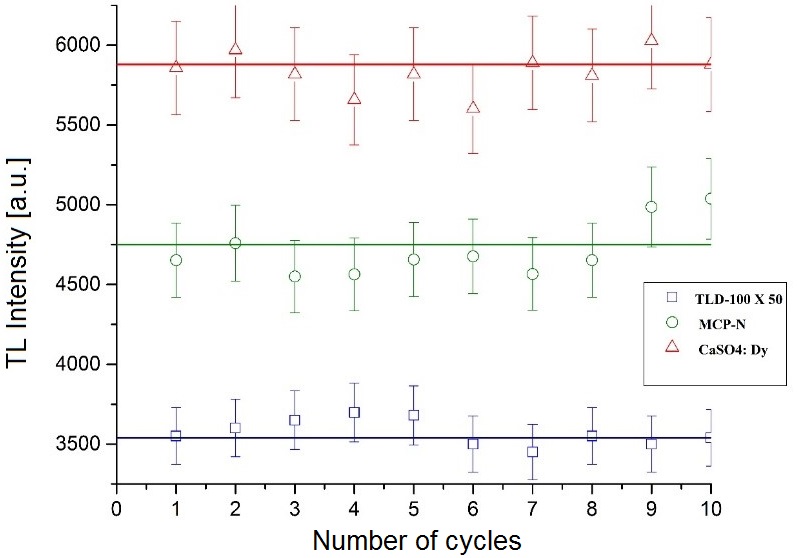}
\caption{Repeatability test for LiF:Mg,Cu,P and CaSO$_4$:Dy compared to the TLD-100 after 10 cycles of irradiation. Note the values of the TLD-100 are multiplied by 50.}
\label{F2}
\end{figure}

The three materials showed values below $7.5$\%, \textit{i.e.}, $3.18$\% for TLD-100, $3.20$\% for CaSO$_4$:Dy and $4.0$\% for the LiF:Mg,Cu,P. In general the three materials showed very good repeatability for low dose X-ray. In Figure \ref{F2}, the values for all materials are presented and include error bar of 5\%.

CaSO$_4$:Dy is about 82\% more sensitivity than TLD-100, while the LiF:Mg,Cu,P was about 66\% higher than the TLD-100. In other investigations \cite{Maia2010}, \cite{Gonzalez2013}, \cite{Lakshmanan2001} have been reported sensitivity factor values about 60\% more than TLD-100 for CaSO$_4$:Dy. The most sensitive is CaSO$_4$:Dy and the second one is LiF:Mg,Cu,P. Among all materials the TLD-100 presented the lowest TL sensitivity.

It is not easy to compare sensitivity values obtained in this study with those from the literature, because this parameter varies significantly with beam energy. Most frequently, sensitivity values are presented for the $^{60}$Co energy, which are not useful for dosimetry purposes in low energy beams. The results obtained showed that, in the diagnostic radiology energy range, the differences in sensitivity among the materials are even more accentuated than for the high energy beam of $^{60}$Co.

Typical TL reproducibility values are between 2\% and 10\% \cite{Campos1983}, \cite{Campos1993},\cite{Oberhofer1981}, \cite{Oliveira2004}. The results of this study were all within the expected range. Moreover, all materials, presented very good performance, with reproducibility values below 4\%.

\begin{figure}[h]
\includegraphics[width=7cm]{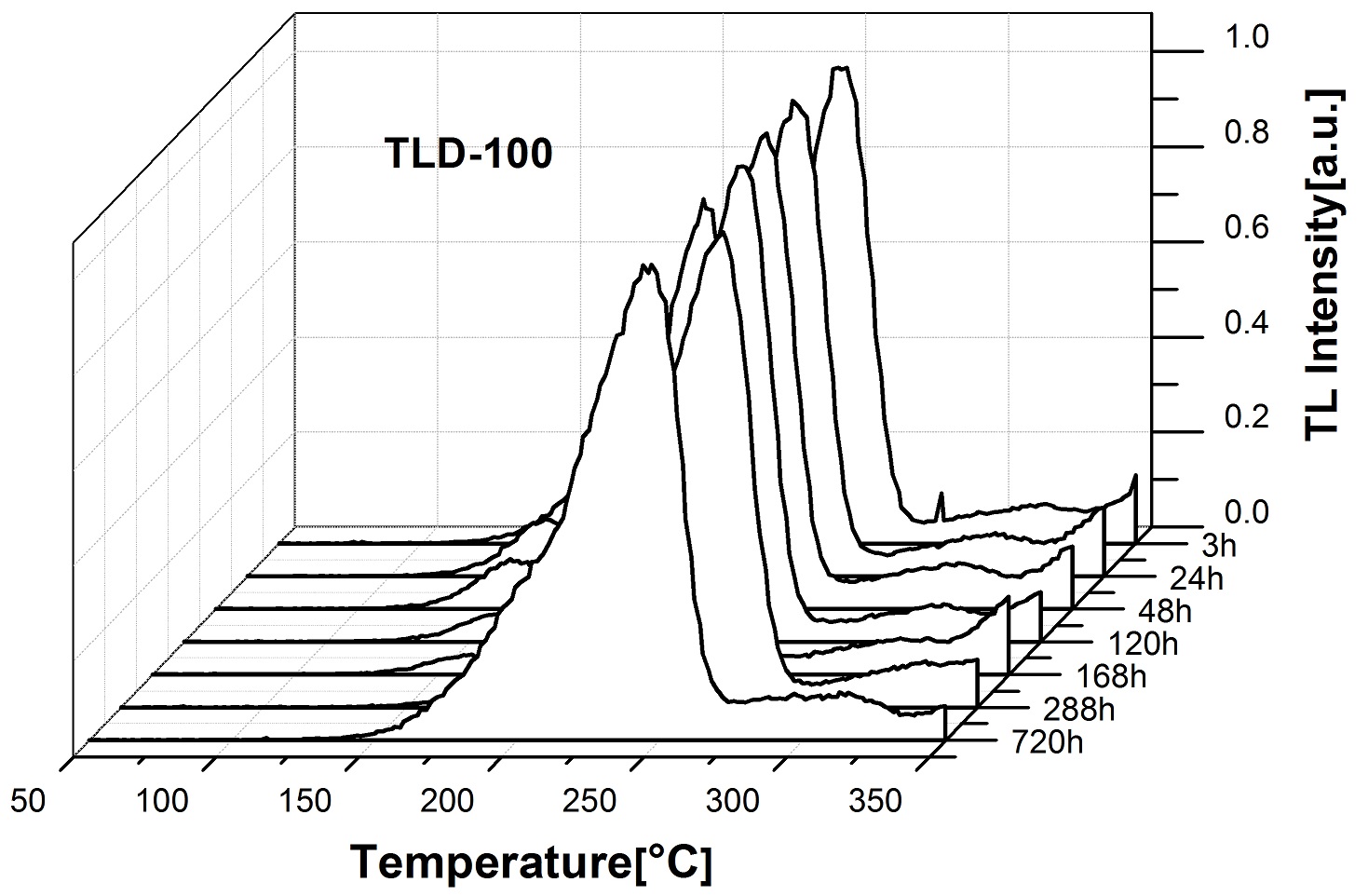}
\caption{Graphical representation of the evolution of the glow curves for TLD-100 from 3 hours (Note 24h) up to 1 month (note 720h) after irradiation.}
\label{F4a}
\end{figure}

From the three materials, LiF:Mg,Cu,P presents less reproducibility, reading up to $240^oC$ was not enough to annealing. Specifically, there is a distribution of traps associated with low intensity peaks located between $270^oC$ and $300^oC$ that are not annealed \cite{Muniz2003}. For this reason, upon reading residual signals remain dependent on their dosimetric history, so after a certain time measurement reproducibility is impoverished.

\begin{figure}[h]
\includegraphics[width=7cm]{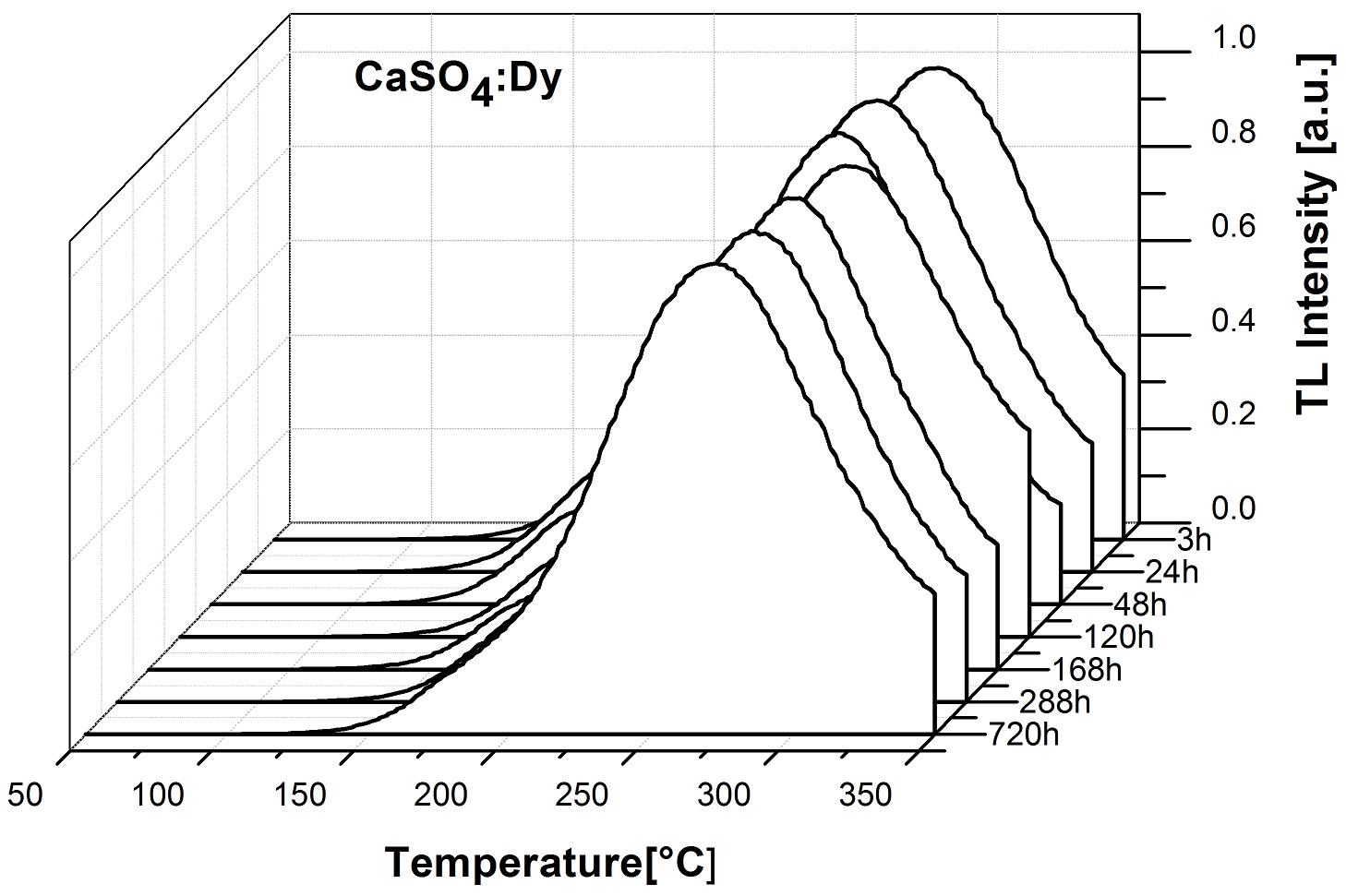}
\caption{Graphical representation of the evolution of the glow curves for CaSO$_4$:Dy from 3 hours (note 24h) up to 1 month (note 720h) after irradiation.}
\label{F4b}
\end{figure}

\begin{figure}[h]
\includegraphics[width=8cm]{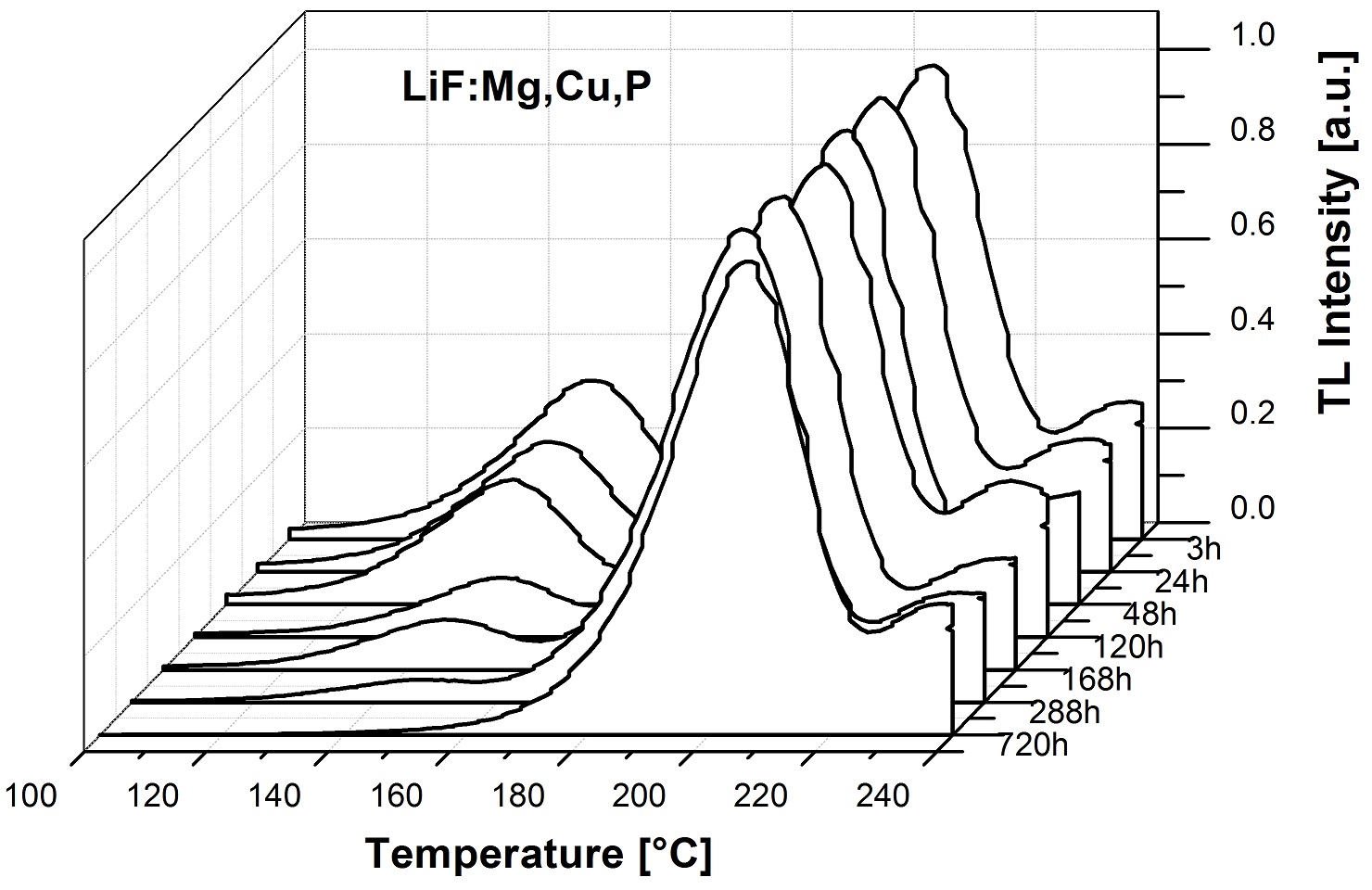}
\caption{Graphical representation of the evolution of the glow curves for LiF:Mg,Cu,P from 3 hours (note 24h) up to 1 month (note 720h) after irradiation.}
\label{F4c}
\end{figure}

In Figures \ref{F4a}, \ref{F4b} and \ref{F4c}, fading of peaks of glow curves for the three materials are shown in a period of one month. It is observed clearly the slight decrease in the intensity of the dosimetric peak for each of the materials, the first peak had a higher fading, completely disappearing at the $288h$ for the TLD-100 and CaSO$_4$:Dy and $720h$ (1 month) for LiF:Mg,Cu,P.

The above mentioned figures (\ref{F4a}, \ref{F4b} and \ref{F4c}) show a small shift to higher temperature are seen in the peak temperature of the main dosimetry peak. 
It is known that a glow peak with kinetic order greater than one (non-first-order TL glow peak) shifts to higher temperatures with decreasing the population of trapping states. Storing the TL dosimeter causes depopulation of trapping states due to fading. Therefore, the TL glow peaks shift to higher temperature with increase in storage time.

\begin{figure}[h]
\includegraphics[width=8cm]{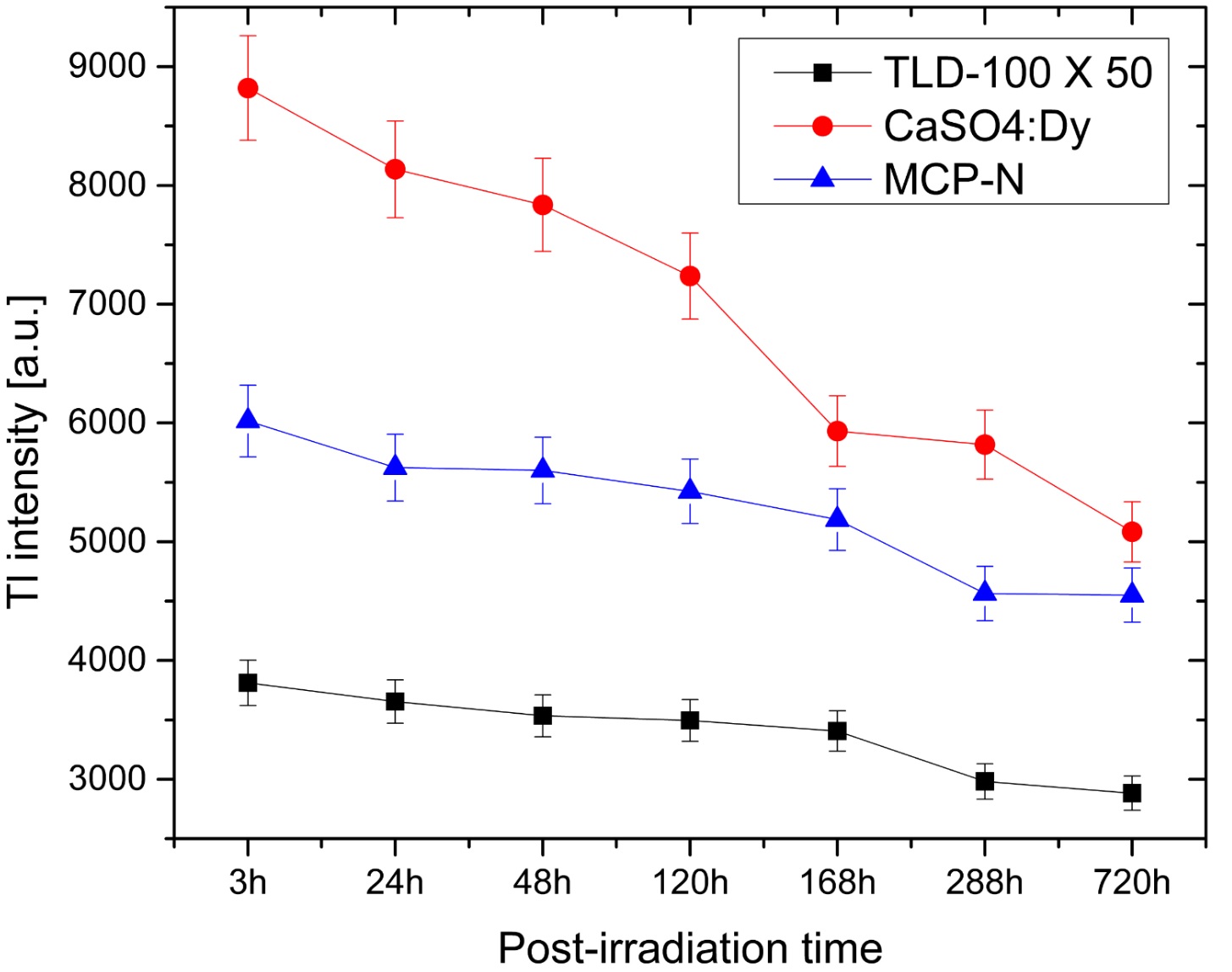}
\caption{Fadings for LiF:Mg,Cu,P from 3 hours (Note 24h) up to 1 month (note 720h) after irradiation.}
\label{FR}
\end{figure}

By analyzing the relative intensities (as shown in Figure \ref{FR}) obtained for each measurement, the TLD-100 showed a slow fading (11\%) for a period of 3 hours to 48 hours, after this the fading was 15\% between 48h and 720h. In the case of CaSO$_4$:Dy the fading was 11\% between 3h and 48h  and 35\% from 48h up to 720h presenting the greatest fading compared to other materials. Finally, LiF:Mg,Cu,P presented a fading of 7\% between 3h and 48h and slow fading (8.8\%) from 48h until 720h post irradiation.

The CaSO$_4$:Dy was the material with a higher fading in the period of one month at room temperature, which is in agreement with that reported in the literature \cite{Chougaonkar2008}, \cite{Yang2002}, in the case TLD-100 showed a high fading as reported by \cite{Carlson2011} and \cite{Davis2003}, while the LiF:Mg,Cu,P dosimeter was experienced slower fading due to loss of some of the initially trapped charges, between irradiation and reading the influence of heat (even at room temperature, thermal fading) or exposure to unwanted light (optical fading). It is further known that the response of LiF:Mg,Cu,P is more stable at ambient temperature than the TLD-100 \cite{McKeever1995}, \cite{Wang1993}.

\section{Conclusions}

In this research, thermoluminescence materials LiF:Mg,Cu, P and CaSO$_4$:Dy were characterized to low doses of X-rays, which correspond to radiological diagnosis by the following dosimetric tests: homogeneity batch reproducibility, sensitive factor, detection threshold, linearity and fading. The experiments were carried out simultaneously with the tests for TLD-100, so that the results are directly comparable.

The materials have a linear behavior for the range of doses studied (CaSO$_4$:Dy and LiF:Mg,Cu,P). TL reading for TLD-100 have a high uncertainty below $4mGy$ so we conclude that is not precise and it has a non-linear behavior in the dose range described above. This findings are very important and should be made available to researchers and medical practitioners that they use TLD-100 dosimeters for low dose measurement in diagnostic radiology.  

The CaSO$_4$:Dy is the material with higher threshold detection with very high sensitivity to low doses X-rays so we suggest its use in environmental and personal dosimetry in diagnostic radiology.

LiF:Mg,Cu,P has high sensitivity and excellent dosimetric characteristics better than the TLD-100, and is the material that show fewer fading in natural light and environmental conditions. We suggest its use for monitoring both environmental and occupational doses in rooms with low doses of radiation.

Finally we conclude that the system formed by the CaSO$_4$:Dy and LiF:Mg,Cu,P is effective for detection of very low doses delivered by X-ray equipment. This system is very useful to physiologists, medical physics, occupationally exposed workers, etc. Our final suggestion is that in every area of health where work with ionizing radiation combined use of both materials becomes necessary to achieve a comprehensive dosimetry monitoring.

\section*{Acknowledgements}
The authors thank SNI-CONACYT Mexico. SDSF acknowledge the student grant by CONACYT and BEIFI grant by IPN. This work was partly supported by SIP20160512, SIP20160522, SIP20150188, COFAA-IPN, EDI-IPN and SNI-CONACYT grants. The authors are indebted to the anonymous referees for their constructive criticism.


\begin{thebibliography}{00}


\bibitem[(DeWerd and Wagner, 1999)]{Dewerd1999} Dewerd, L.A., Wagner, L.K., (1999). Characteristics of radiation detectors for diagnostic radiology, Appl. Radiat. Isot. 50, 125–136.

\bibitem[(Guimaraes \textit{et al.}, 2003)]{Guimaraes2003} 
Guimaraes, C.C., Okuno, E., 2003. Blind performance testing of personal and environmental dosimeters based on TLD-100 and natural CaF$_2$:NaCl. Radiat. Meas. 37, 127-132. 

\bibitem[(IAEA, 1995)]{IAEA1995} 
IAEA: International Atomic Energy Agency, 1995. Radiation doses in diagnostic radiology and methods of dose reduction IAEA-TECDOC-796 (Vienna: IAEA).

\bibitem[(Horowitz, 1984)]{Horowitz1984} 
Horowitz, Y.S., 1984. Thermoluminescence and Thermoluminescent Dosimetry, 3 vols, CRC Press, Boca Raton, FL, USA.

\bibitem[(DeWerd, 1979)]{DeWerd1979} DeWerd L A (1979). “Application of thermally induced luminescence” Thermally Stimulated Relaxation in Solids. Editor: P. Braunlich Berliir Springer-Verlag.

\bibitem[(Oberhofer and Scharmann, 1981)]{Oberhofer1981}
Oberhofer, M., Scharmann, A., (1981). Applied Thermoluminescence Dosimetry. Adam Hilger, Bristol.

\bibitem[(McKeever, 1985)]{McKeever1985} McKeever S.W.S. (1985). Thermoluminiscence of Solids, World Scientific, First edition, New Jersey.

\bibitem[(Bilski \textit{et al.}, 1995)]{Bilski1995}
Bilski,P. Olko,P. Burgkhardt,B. Piesch,E., (1995). Ultra-Thin LiF:Mg,Cu,P Detectors for Beta Dosimetry. Radiat. Meas. 24, 439-443.

\bibitem[(Budzanowski \textit{et al.}, 2007)]{Budzanowski2007} 
Budzanowski M., Bilski P., Olko P., Ryba E., Perle S., Majewski M., (2007). Dosimetric properties of new cards with high-sensitivity MCP-N (LiF:Mg,Cu,P) detectors for Harshaw automatic reader. Radiat. Prot. Dosim. Vol. 125(4), 251-253.

\bibitem[(Sukis, 1971)]{Sukis1971} Sukis, D. R. (1971). Thermoluminescent properties of CaF2: Dy TLD's. Nuclear Science, IEEE Transactions on, 18(6), 185-189.

\bibitem[(Furetta and Lee, 1983)]{Furetta1983} Furetta, C., \& Lee, Y. K. (1983). Annealing and fading properties of CaF$_2$: Tm (TLD-300). Radiation protection dosimetry, 5(1), 57-63.

\bibitem[(Becker, 1972)]{Becker1972} Becker, K. (1972). Environmental monitoring with TLD. Nuclear Instruments and Methods, 104(2), 405-407.

\bibitem[(Vohra \textit{et al.}, 1980)]{Vohra1980} Vohra, K. G., Bhatt, R. C., Chandra, B., Pradhan, A. S., Lakshmanan, A. R., \& Shastry, S. S. (1980). A Personnel Dosimeter TLD Badge Based on CaSO$_4$:Dy Teflon TLD Discs. Health physics, 38(2), 193-197.

\bibitem[(Benkrid \textit{et al.}, 1992)]{Benkrid1992} Benkrid, M., Mebhah, D., Djeffal, S., \& Allalou, A. (1992). Environmental gamma radiation monitoring by means of TLD and ionisation chamber. Radiation Protection Dosimetry, 45(1-4), 77-80.

\bibitem[(Takale \textit{et al.}, 2014)]{Takale2014} Takale, R. A., Sahu, S. K., Swarnkar, M., Shetty, P. G., \& Pandit, G. G. (2014). Optimizations of teflon embedded CaSO4: Dy based TLD for environmental monitoring applications. Journal of Radioanalytical and Nuclear Chemistry, 302(3), 1405-1411.

\bibitem[(Furetta and Weng, 1998)]{Furetta1998} Furetta C., Weng P., (1998). Operational Thermoluminescence Dosimetry. World Scientific. First Edition. Singapore.

\bibitem[(Hirning, 1992)]{Hirning1992} 
Hirning, C. R., 1992. Detection and determination limits for thermoluminiscence dosimetry. Health Phys. 62(3), 223-7.

\bibitem[(Livingstone \textit{et al.}, 2009)]{Livingstone2009} 
Livingstone, J., Horowitz, Y. S., Oster, L., Datz, H., Lerch, M., Rosenfeld, A. Horowitz, A., 2009. Experimental investigation of the $100 keV$ x-ray dose response of high-temperature thermoluminescent in LiF:Mg,Ti (TLD-100): theoretical interpretation using the unified interaction model. Radiat. Prot. Dosim., 138(4), 320–333.

\bibitem[(Maia and Caldas, 2010)]{Maia2010} 
Maia A.F., Caldas L.V., 2010. Response of TL materials to diagnostic radiology X radiation beams. Appl Radiat Isot. 68(4-5), 780-3.

\bibitem[(Cameron \textit{et al.}, 1968)]{Cameron1968} 
Cameron, J. R., Suntharalingam, N. Kenney, G. N., (1968). Thermoluminescent Dosimetry. Madison: University of Wisconsin Press.

\bibitem[(Watson, 1970)]{Watson1970} 
Watson, C. R., 1970. Linearity of TLD response curves. Health Phys., 18, 168–169.

\bibitem[(Gonz\'alez \textit{et al.}, 2013)]{Gonzalez2013} 
Gonz\'alez P.R, Cruz-Zaragoza E., Furetta C., Azor\'in J., Alc\'antara B.C., 2013. Effect of thermal treatment on TL response of CaSO$_4$:Dy obtained using a new preparation method. Appl Radiat Isot. 75, 58-63.

\bibitem[(Lakshmanan, 2001)]{Lakshmanan2001}
Lakshmanan, A.R. (2001), A New High Sensitive CaSO$_4$:Dy Thermostimulated Luminescence Phosphor. Physica Status Solidi(a), 186(1), 153-166.

\bibitem[(Campos, 1983)]{Campos1983}
Campos, L.L., (1983). Preparation of CaSO4:Dy TL single crystals. J. Lumin. 28 (4), 481–483.

\bibitem[(Campos, 1993)]{Campos1993}
Campos, L.L., (1993). Graphite mixed CaSO$_4$:Dy TL dosemeters for beta radiation dosimetry. Radiat. Prot. Dosim. 48 (2), 205–207.

\bibitem[(Oliveira and Caldas, 2004)]{Oliveira2004}
Oliveira, M.L., Caldas, L.V.E., (2004). Performance of different thermoluminescence dosemeters in $^{90}$Sr+$^90$Y radiation fields. Radiat. Prot. Dosim. 111 (1), 17–20.

\bibitem[(Mu\~niz, 2003)]{Muniz2003} 
Mu\~niz, J.L., (2003). Métodos experimentales de dosimetría postal para el control de calidad en radioterapia basados en LiF:Mg, Ti (TLD-100) y LiF:Mg, Cu P (GR-200) : aplicación de métodos numéricos al análisis de las curvas de termoluminiscencia. (Experimental Methods postal dose for quality control based radiotherapy LiF: Mg, Ti (TLD-100) and LiF: Mg, Cu P (GR-200): application of numerical methods to the analysis of the curves thermoluminescence) PhD. Dissertation. Universidad Complutense de Madrid, Espa\~na.

\bibitem[(Chougaonkar \textit{et al.}, 2008)]{Chougaonkar2008} 
Chougaonkar, M. P., Takle R. A., Mayya Y. S., Puranik V. D., Kushwaha H. S., (2008). Performance Characteristics of Newly Modified CaSO$_4$:Dy based Indigenous Thermoluminescent Dosimeters for Environmental Radiation Monitoring. Jour.Nucl. Sci.Tech. 45(5), 610-613.

\bibitem[(Yang \textit{et al.}, 2002)]{Yang2002} 
Yang J.S., Kim D.Y., Kim J.L., Chang S.Y., Nam Y.M., Park J.W., 2002. Thermoluminescence characteristics of teflon embedded CaSO$_4$:Dy TLD. Rad. Prot. Dosim. 100(1-4), 337-40.

\bibitem[(Carlson \textit{et al.}, 2011)]{Carlson2011} 
Carlson Tedgren A., Hedman A., Grindborg J.E., Carlson G.A., (2011). Response of LiF: Mg,Ti thermoluminescent dosimeters at photon energies relevant to the dosimetry of brachytherapy ($<1 MeV$). Med. Phys., 38, 5539–5550.

\bibitem[(Davis \textit{et al.}, 2003)]{Davis2003} 
Davis S.D., Ross C., Mobit P., Van Der Zwan L., Chase W., Shortt K., (2003). The response of LiF thermoluminescence dosemeters to photon beams in the energy range from $30 kV$ X rays to $^{60}$Co gamma rays. Radiat. Prot. Dosim., 106, 33–43.

\bibitem[(McKeever \textit{et al.}, 1995)]{McKeever1995} 
McKeever S.W.S., Moscovitch M., Townsend P.D., (1995). Thermoluminescence Dosimetry Materials: Properties and Uses. Nuclear Technology Publishing, Ashford, UK.

\bibitem[(Wang \textit{et al.}, 1993)]{Wang1993} 
Wang J, Wu F, Zhao J, Wu F, Zhu J, Li Y, (1993). A TLD reader based on single chip microprocessor, Nuclear Tracks and Radiation Measurements, 22(1–4), 893-895.





\end{thebibliography}
\end{document}